\documentclass[conference]{IEEEtran}
\IEEEoverridecommandlockouts
\usepackage{cite}
\usepackage{amsmath,amssymb,amsfonts}
\usepackage{algorithmic}
\usepackage{graphicx}
\usepackage{textcomp}
\usepackage{xcolor}
\usepackage{hyperref}

\hypersetup{
    colorlinks=true, 
    linkcolor=black,  
    citecolor=black, 
    filecolor=black, 
    urlcolor=black    
}

\def\BibTeX{{\rm B\kern-.05em{\sc i\kern-.025em b}\kern-.08em
    T\kern-.1667em\lower.7ex\hbox{E}\kern-.125emX}}
\begin{document}

\title{
Iterative Matrix Product State Simulation for Scalable Grover’s Algorithm
\thanks{*Corresponding Author: 2503001@narlabs.org.tw\\}
}

\author{
\IEEEauthorblockN{
    Mei Ian Sam \IEEEauthorrefmark{1},
    Tzu-Ling Kuo \IEEEauthorrefmark{2}\IEEEauthorrefmark{3},
    Tai-Yue Li \IEEEauthorrefmark{3}*,
}
\IEEEauthorblockA{\IEEEauthorrefmark{1}Department of Physics, National Tsing Hua University, Hsinchu, Taiwan}
\IEEEauthorblockA{\IEEEauthorrefmark{2}Undergraduate Program in Intelligent Computing and Big Data, Chung Yuan Christian University, Taoyuan, Taiwan}
\IEEEauthorblockA{\IEEEauthorrefmark{3}National Center for High-performance Computing, National Institutes of Applied Research, Hsinchu, Taiwan}
}

\maketitle
\begin{abstract} 
Grover’s algorithm is a cornerstone of quantum search algorithm, offering quadratic speedup for unstructured problems. However, limited qubit counts and noise in today’s noisy intermediate-scale quantum (NISQ) devices hinder large-scale hardware validation, making efficient classical simulation essential for algorithm development and hardware assessment. We present an iterative Grover simulation framework based on matrix product states (MPS) to efficiently simulate large-scale Grover's algorithm. Within the NVIDIA CUDA-Q environment, we compare iterative and common (non-iterative) Grover's circuits across statevector and MPS backends. On the MPS backend at 29 qubits, the iterative Grover's circuit runs about 15× faster than the common (non-iterative) Grover's circuit, and about 3-4× faster than the statevector backend. In sampling experiments, Grover's circuits demonstrate strong low-shot stability: as the qubit number increases beyond 13, a single-shot measurement still closely mirrors the results from 4,096 shots, indicating reliable estimates with minimal sampling and significant potential to cut measurement costs. Overall, an iterative MPS design delivers speed and scalability for Grover's circuit simulation, enabling practical large-scale implementations.
\end{abstract}


\section{Introduction} 
Grover’s algorithm~\cite{grover1996fast} is a fundamental quantum algorithm that achieves a quadratic speedup for unstructured search, reducing the query complexity from $O(N)$ to $O(\sqrt{N})$, where $N$ denotes the size of the unstructured search space ($N = 2^n$, $n$ is the number of qubits). Bennett et al.~\cite{bennett1997strengths} established the $\Omega(\sqrt{N})$ lower bound for any quantum search, proving Grover’s optimality within quantum mechanics. Boyer et al.~\cite{boyer1998tight} further derived the exact query complexity and success probability, confirming that Grover’s amplitude amplification achieves this bound precisely. Owing to its optimal efficiency, Grover’s method has been extended beyond database search to combinatorial optimization~\cite{durr1996quantum, baritompa2005grover, gilliam2021grover}, constraint satisfaction~\cite{cerf2000nested}, and quantum machine learning tasks including classification~\cite{du2021grover, muser2024provable, zhao2022binary, tai2022quantum, an2025quantum, chen2024validating, hsu2025quantum, hsuKernel2025quantum}, pattern recognition~\cite{tezuka2022grover}, and feature selection~\cite{laius2023quantum}. 

Despite its theoretical elegance, practical realization on current NISQ hardware remains difficult~\cite{preskill2018quantum}. The circuit depth and gate numbers increase rapidly with the number of qubits $n$, amplifying error sensitivity and decoherence, while the required coherence times far exceed current hardware capabilities. Hence, efficient simulation frameworks that capture Grover’s iterative dynamics without constructing deep circuits are crucial for its practical study.

In this work, we introduce an \textit{Iterative Grover's Circuit} that reproduces the full Grover process without explicitly generating all $O(\sqrt{N})$ layers, thus achieving a substantial reduction in memory requirements. Our approach iteratively updates the quantum state by repeatedly applying a single Grover gate, enabling scalable simulations on both state-vector and MPS backends. We evaluate runtime performance and further analyze the effects of floating-point precision and measurement shots on simulation accuracy.

For systems with more than $27$ qubits, our iterative MPS simulation outperforms state-vector methods in runtime (and memory). In particular, for $n=29$, the iterative MPS achieves over a $3$× speedup compared to state-vector backend, and more than a $15$× improvement over common MPS implementations. Accurate amplitude reconstruction is achieved with floating point 64 (FP64) precision and few measurement shots (e.g. 1 or 8 shots) even for $n \ge 13$, substantially lowering computational cost. These results demonstrate that the proposed iterative scheme enables efficient and precise large-scale Grover simulation beyond current hardware limitations.

\section{Grover's Algorithm} 
Grover’s algorithm demonstrates that quantum computation can solve unstructured search problems in $O(\sqrt{N})$ time, whereas any classical approach requires $O(N)$ queries. It repeatedly applies the Grover gate $G$ -- a unitary rotation in a two-dimensional subspace spanned by the marked and unmarked components of the quantum state.

Let $N=2^n$ denote the size of the search space ($n$ is the number of qubits) and $M$ the number of marked items. 
Define
\begin{equation}
|w\rangle = \frac{1}{\sqrt{M}}\!\sum_{x \in \text{marked}}\! |x\rangle, 
\qquad
|r\rangle = \frac{1}{\sqrt{N-M}}\!\sum_{x \notin \text{marked}}\! |x\rangle,
\end{equation}
which form an orthonormal basis for the relevant subspace. 

The uniform initial state is
\begin{equation}
|s\rangle = \frac{1}{\sqrt{N}}\!\sum_{x=0}^{N-1}\!|x\rangle
          = \sin\theta\,|w\rangle + \cos\theta\,|r\rangle
\end{equation}

Each Grover iteration is given by
\begin{equation}
G = D O = (I-2|s\rangle\langle s|)(I - 2|r\rangle\langle r|)
\end{equation}

Here, $O = I - 2|r\rangle\langle r|$ is the \textit{Oracle reflection}, which inverts the component of the state along the unmarked subspace spanned by $|r\rangle$. Geometrically, it acts as a mirror reflection along the direction of $|r\rangle$. 

The operator $D = I - 2|s\rangle\langle s|$, known as the \textit{diffusion operator} performs a reflection along the direction of $|s\rangle$ geometrically. Specifically, it can decomposed as:
\begin{equation}
D = H^{\otimes n} (2|0\rangle\langle 0| - I) H^{\otimes n}
   = H^{\otimes n} R_0 H^{\otimes n}
\end{equation}
where $H^{\otimes n}$ denotes the $n$-qubit Hadamard transform and $R_0$ is the reflection about the $|0\rangle$ state.

Together, these two reflections compose the Grover operator $G$, which performs a rotation by $2\theta$ in the two-dimensional subspace $\mathrm{span}\{|w\rangle, |r\rangle\}$.

For $k$ iterations, state $|s\rangle$ becomes:
\begin{equation}
G^k |s\rangle = \sin((2k+1)\theta)|w\rangle + \cos((2k+1)\theta)|r\rangle
\end{equation}
and the success probability for searching marked items is 
\begin{equation}
P(k) = \sin^2((2k+1)\theta),
\end{equation}
which peaks when $(2k+1)\theta = \pi/2$, giving the optimal iteration number
\begin{equation}
k_{\text{opt}} = \frac{\pi}{4\theta} - \frac{1}{2}
\end{equation}
Substituting the small-angle approximation $\theta\approx\sqrt{M/N}$ yields
$k_{\text{opt}} \approx \frac{\pi}{4}\sqrt{N/M} - \frac{1}{2}$.

In our simulation, marked item $M$ is set to 1, and we set the interation number $k$ to
\begin{equation}
    k = \text{round}(\frac{\pi}{4}\sqrt{N})
\end{equation}

\begin{figure}[!t]
\centering
\includegraphics[scale=0.25]{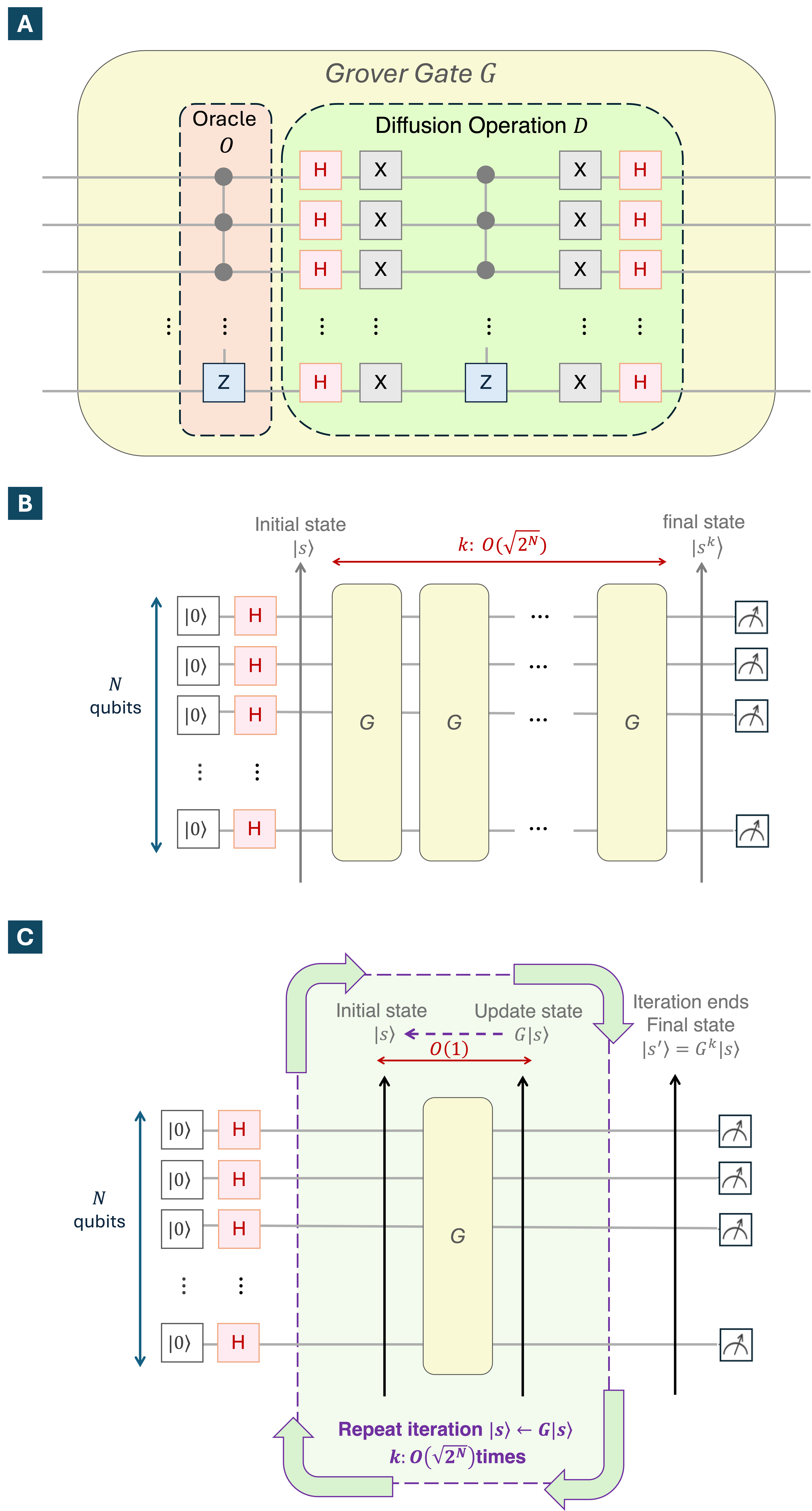}
\caption{\textbf{Structure and comparison of Grover’s quantum circuits.}
(A) Internal composition of the Grover gate $G$, consisting of the oracle $O$ and the diffusion operator $D$. In the circuit, filled circles connected by vertical lines represent multi-controlled operations while the letters denote standard quantum gates: $H$ for Hadamard, $X$ for Pauli-X, and $Z$ for Pauli-Z gates. (B) \textit{Common Grover’s Quantum Circuit:} the standard implementation repeats $G$ for $O(\sqrt{2^n})$ iterations, leading to a very deep quantum circuit as the number of qubits increases. (C) \textit{Iteration Grover’s Quantum Circuit:} the proposed iterative approach updates the state vector $|s\rangle \leftarrow G|s\rangle$ at each step, avoiding the multi-layer circuit construction.} 
\label{fig:grover-circuits}
\end{figure}

\section{Methodology} 
\subsection{Iterative Grover's Quantum Circuits} 
Following the mathematical formulation presented in the previous section, both the oracle $O$ and diffusion operators $D$ in the Grover gate $G$ as illustrated in Fig.~\ref{fig:grover-circuits}(A). Conventionally, the complete Grover algorithm is implemented by repeatedly applying $G$ for $k$ iterations as illustrated in Fig.~\ref{fig:grover-circuits}(B). This approach constructs a deep quantum circuit consisting of many stacked Grover layers, which rapidly increases circuit depth and memory requirements as the number of qubits grows. Although such a structure is conceptually straightforward, it becomes computationally expensive for large-scale simulations and is beyond the limit of current quantum hardware. We will call this structure \textit{Common method} in the following sections.

To overcome this issue, we propose an \textit{Iterative Grover Circuit} Structure, as illustrated in Fig.~\ref{fig:grover-circuits}(C). 
Instead of explicitly generating the entire deep circuit, the algorithm updates the quantum state in each iteration by reapplying a single Grover gate $G$ to the current state. This iterative state-update approach is mathematically equivalent to the conventional implementation, but it avoids the need to construct or store a large layered circuit. We will call this structure \textit{Iterative method} in the following sections.

\subsection{Simulation Backend} 
To evaluate the performance and scalability of Grover’s algorithm with increasing qubit numbers, we employ two quantum simulation backends: the state-vector and matrix product state (MPS). 

\subsubsection{State-vector Simulation}
The state-vector backend represents an $n$-qubit system as a complex vector of size $2^n$, with each gate modeled as a dense matrix acting on this vector. As circuit depth increases, repeated matrix multiplications lead to exponential growth in both runtime and memory, making simulation impractical beyond roughly 30–35 qubits on classical hardware. Despite this cost, state-vector simulation yields numerically exact results and thus serves as the reference for validating our optimized Grover implementation.

\subsubsection{Matrix Product State (MPS) Simulation}
The MPS backend approximates the quantum state using a chain of low-rank tensors connected by a bond dimension $\chi$, truncating small singular values during tensor updates ~\cite{perez2006matrix}. This representation reduces memory scaling from $O(2^n)$ to $O(n\chi^2)$ and allows a tunable trade-off between accuracy and efficiency. After each gate, tensors are locally updated, and truncation is applied if the dimension exceeds a user-defined $\chi_{\max}$ (set to 64 in this study). Only dominant singular values are retained, defining the actual bond dimension $\chi$ used during simulation.


\begin{figure}[!t]
\centering
\includegraphics[scale=0.41]{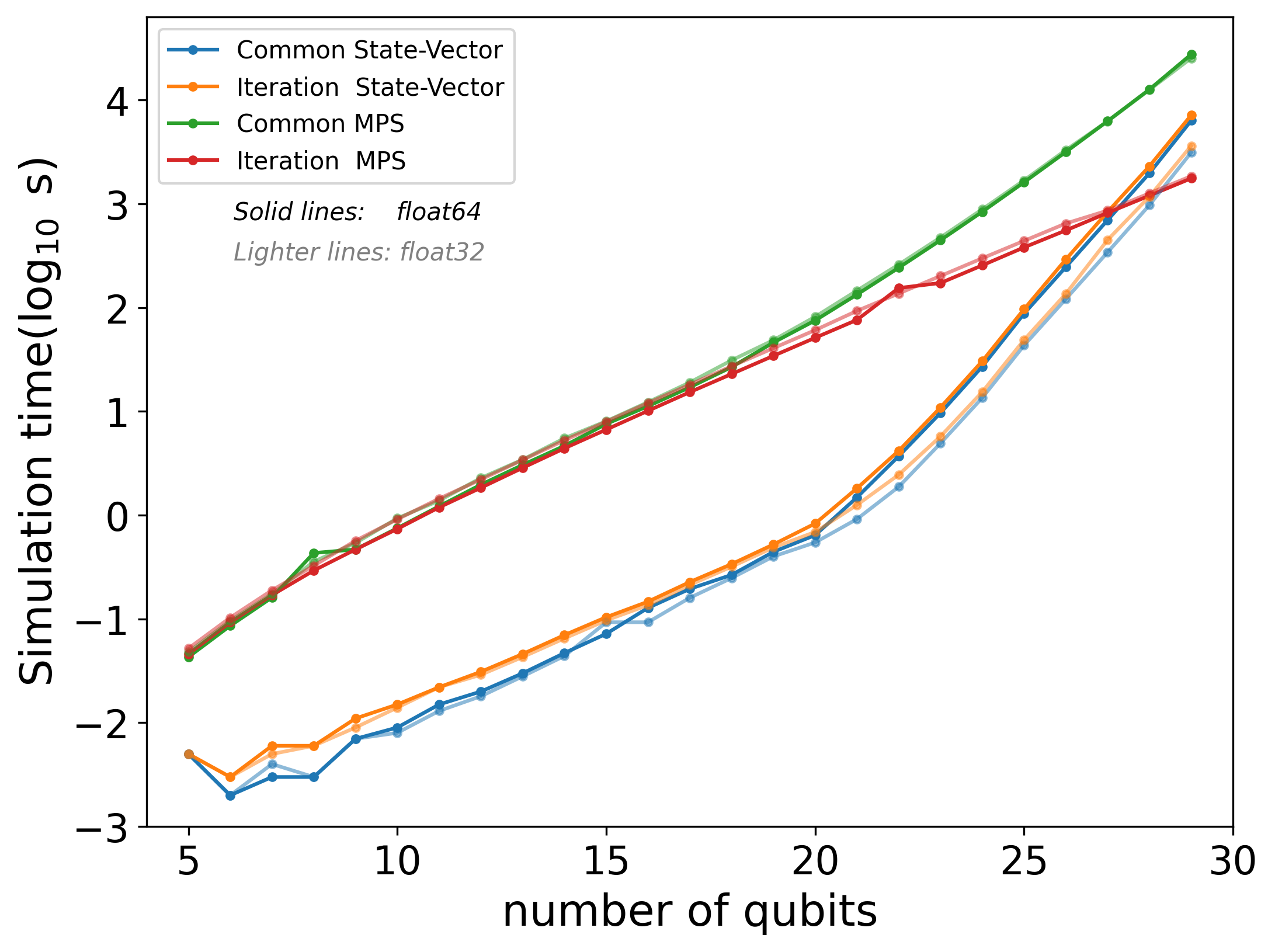}
\caption{\textbf{Runtime of Grover’s algorithm versus the number of qubits for different methods,  simulation backends and precisions.} At small system sizes, the proposed Iterative method and the Common implementation show similar performance, and the state-vector backend runs faster than MPS. As the number of qubits increases, the runtime of state-vector simulation grows most rapidly, followed by Common (MPS), whereas Iterative (MPS) scales the slowest. Beyond 27 qubits, the MPS Iterative method outperforms state-vector simulation, and the observed trend suggests a crossing point between Common (MPS) and state-vector performance at even larger scales.}
\label{fig:grover-scaling}
\end{figure}

\begin{figure*}[!t]
\centering
\includegraphics[scale=0.4]{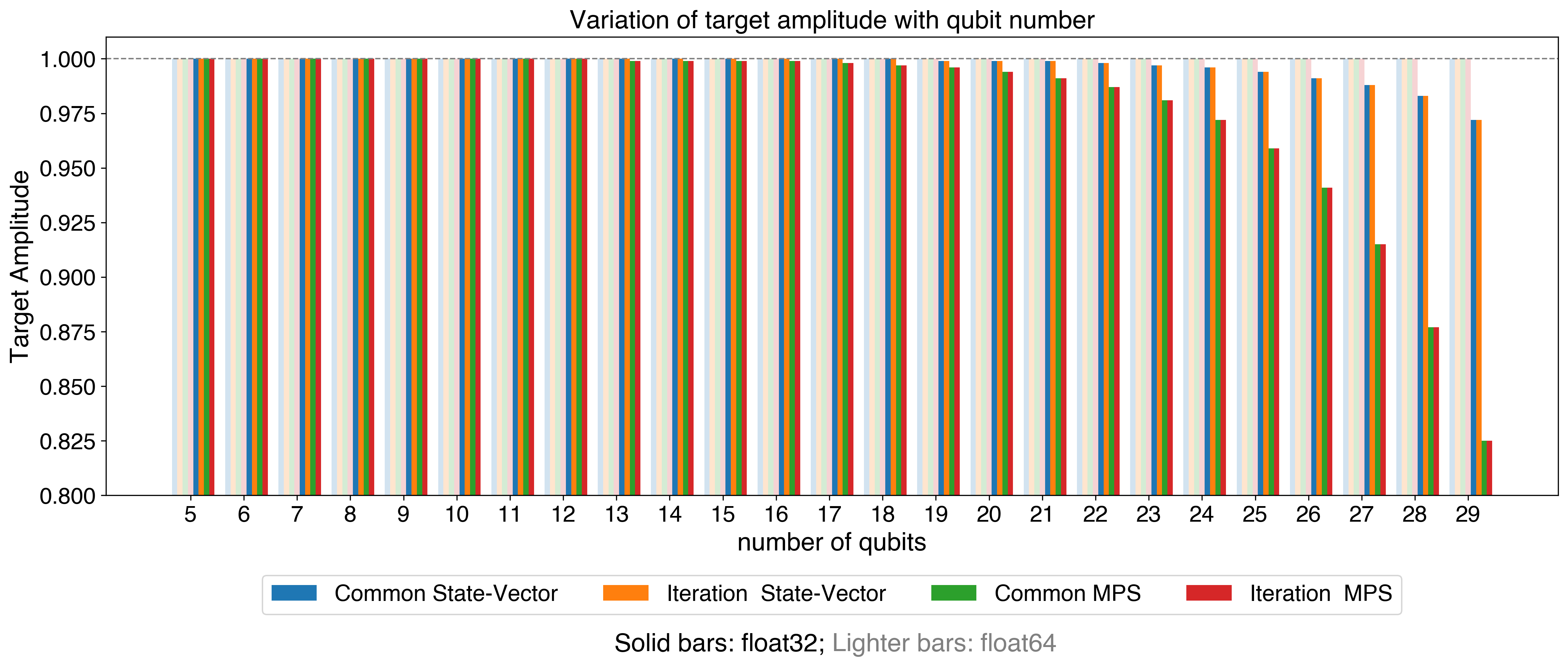}
\caption{\textbf{Target amplitude of Grover’s algorithm obtained from different methods, backends and precisions.} Float64 state-vector simulations (lighter bars) serve as the reference and yield identical amplitudes of 1.0 across all setups. When using float32 (solid bars), the target amplitude remains consistent between the Common and Iterative implementations for small systems, but deviations appear as the qubit number increases. MPS amplitudes start to decrease at $n=18$ and drop below 0.9 at $n=28$, whereas state-vector amplitudes begin to decrease only at $n=23$ and remain above 0.98.
}
\label{fig:grover-target-amplitude}
\end{figure*}

\section{Result}
\subsubsection{Simulation Time Benchmark}
Figure\ref{fig:grover-scaling} presents the simulation time scaling of Grover’s algorithm for different structures, simulation backends and floating-point precisions. Let's focus on the solid lines, the results of FP64 precision, first. For small qubit systems ($n\leq 20$), the proposed iterative method and the conventional common construction exhibit nearly identical performance. In this regime, the state-vector backend significantly outperforms the MPS backend, the state vector backend runs up to $68.97$× faster than the MPS backend for Iterative method at $n = 16$, which represents the largest performance gap observed when the state-vector backend outperforms the Iterative (MPS) implementation.

As the number of qubits increases, the state-vector runtime rises most rapidly, followed by Common (MPS), whereas Iterative (MPS) exhibits the slowest growth. When the qubit number exceeds 27, Iterative (MPS) surpasses all state-vector counterparts in efficiency. Specifically, at $n=29$, it achieves speedups of $4.06$× over the Iterative (State vector) simulation, $3.61$× over the Common (State vector) method, and $15.66$× over the Common (MPS) implementation. The observed slopes further suggest that Common (MPS) and state-vector curves would intersect at larger system sizes ($n>30)$, highlighting the intrinsic efficiency of the Iterative method in reducing computational scaling and memory overhead.

A comparison between 32-bit and 64-bit floating-point precisions shows that the overall scaling trend is nearly identical across all backends. Lower precision offers only a slight runtime gain for the state-vector simulation at large qubit counts, while the growth rate with respect to $n$ remains unchanged. Even under 32-bit precision, the iterative MPS backend surpasses the state-vector simulation beyond $n \approx 28$, confirming its superior scalability and memory efficiency for large-scale Grover simulations.

\begin{figure*}[h]
\centering
\includegraphics[scale=0.4]{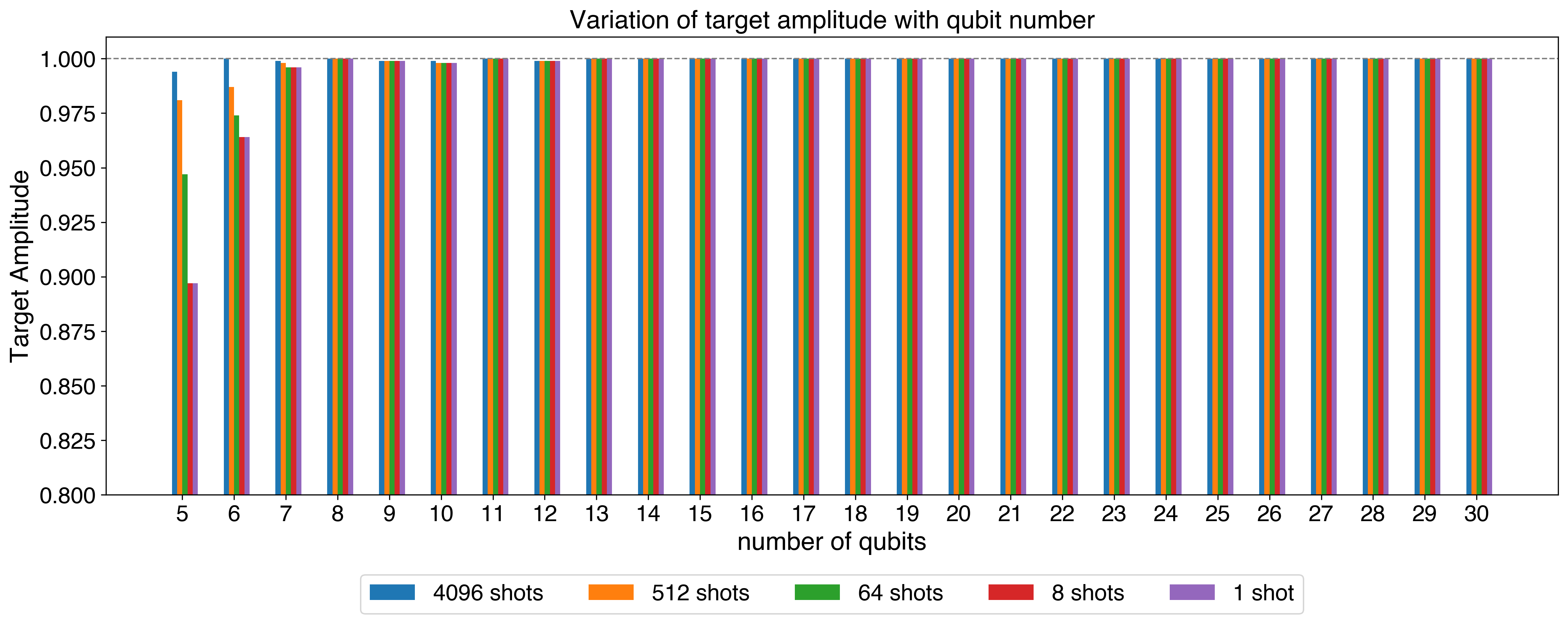}
\caption{\textbf{Effect of measurement shot number on the target amplitude in Grover’s algorithm, obtained using the MPS Iterative simulation with FP64 precision.} For small qubit systems ($n \le 7$), Low-shot measurements (1 or 8 shots) lead to visible amplitude fluctuations, whereas for larger qubit numbers $(n \geq 13)$ the results from all shot numbers converge to $1.0$.}
\label{fig:grover-target-amplitude-shots}
\end{figure*}

\subsubsection{Amplitude Fidelity Benchmark}
To evaluate the numerical accuracy of our method, we measured and compared the target-qubit amplitude obtained from various configurations. Results are presented in Figure \ref{fig:grover-target-amplitude}. Using the FP64 state-vector simulation as our reference, we observe an ideal amplitude of $1.0$ in every case, confirming that both the Common and Iterative methods—across both backends—perfectly reproduce Grover’s amplification. Under FP32, both implementations maintain identical amplitudes at small qubit counts, but precision errors grow as the number of qubits increases. The MPS backend begins to show a slight amplitude reduction at $n=18$ (amplitue $=0.997$) while the state-vector amplitude remains at 1.0 up to $n=23$, where it also starts to decrease to $0.997$. The amplitude drop in MPS becomes significant at larger system sizes, reaching $0.877$ at $n=28$ whereas the state-vector simulation still maintains above $0.98$. This indicates that for large-scale systems $n \geq 28$, where MPS iterative already outpaces the state-vector—double precision (FP64) is necessary to maintain numerical correctness and amplitude fidelity.

\subsubsection{Low-Shot Sampling Benchmark}
We examine how the target amplitude varies with different numbers of measurement shots using the MPS Iterative simulation backend under FP64 precision. The number of shots is gradually reduced from $4096$ down to $1$ to evaluate how limited sampling affects the accuracy of the measured amplitude. As shown in Figure\ref{fig:grover-target-amplitude-shots}, the amplitude decreases noticeably when very few shots are used, particularly for small qubit systems ($n \leq 6$), where statistical undersampling dominates. For example, at $n=5$ the measured amplitude drops from $0.994$ ($4096$ shots) to below $0.897$ ($1$ shot), corresponding to a relative error of $9.76\%$.

However, as the number of qubits grows, this effect rapidly diminishes, and the amplitudes converge to $1.0$ across all shot configurations after $n=13$. This trend indicates that, for large-scale MPS Iterative simulations, Grover’s amplitude amplification is inherently robust against sampling noise, and even a small number of shots (e.g., 1 or 8) is sufficient to reproduce the expected target amplitude. It significantly reduces computational cost and sampling time while maintaining accuracy, enabling scalable studies of large-qubit systems.

\section{Conclusion}
This study presents an optimized simulation framework for Grover’s algorithm based on the Iterative method with an MPS backend, achieving improved scalability and high accuracy. By reusing a single Grover operator instead of constructing deep composite circuits, the proposed approach significantly reduces memory and runtime while preserving identical algorithmic behavior. When the system size exceeds $27$ qubits, the Iterative MPS simulation under FP64 precision outperforms the Common state-vector method, demonstrating clear advantages in efficiency and scalability. Moreover, the target amplitude remains robust under few-shot measurements for large systems ($n \ge 13$), implying shorter execution cycles and reduced measurement-induced errors on practical quantum devices. Overall, the iterative MPS approach provides a scalable, accurate, and resource-efficient framework for simulating and benchmarking Grover’s algorithm on classical hardware.

\section*{Acknowledgment} 
The successful completion of this research was made possible by the academic resources and advanced research infrastructure provided by the National Center for High-Performance Computing, National Institutes of Applied Research (NIAR), Taiwan. We gratefully acknowledge their invaluable support.
\bibliographystyle{siamurl}
\bibliography{references}

\end{document}